\documentclass[pdflatex,sn-mathphys]{sn-jnl}% Math and Physical Sciences Reference Style
\jyear{2022}%

\theoremstyle{thmstyleone}%
\theoremstyle{thmstyletwo}%

\theoremstyle{thmstylethree}%

\begin{document}

\title[All-passive Microwave-Diode Nonreciprocal Metasurface]{All-passive Microwave-Diode Nonreciprocal Metasurface}

\author*[1]{\fnm{Xiaozhen} \sur{Yang}}\email{xiy003@eng.ucsd.edu}

\author[2]{\fnm{Erda} \sur{Wen}}\email{ewen@eng.ucsd.edu}
\equalcont{These authors contributed equally to this work.}

\author[1]{\fnm{Daniel} \sur{Sievenpiper}}\email{dsievenpiper@eng.ucsd.edu}
\equalcont{These authors contributed equally to this work.}

\affil*[1]{\orgdiv{Electrical and Computer Engineering Department}, \orgname{University of California, San Diego}, \orgaddress{\street{9500 Gilman Dr.}, \city{La Jolla}, \postcode{92093}, \state{California}, \country{USA}}}

\abstract{Breaking reciprocity in the microwave frequency range will have important implications for modern electronic systems. Since it usually involves bulky biasing magnets or complex spatial-temporal modulations, exploring a lightweight, all-passive approach becomes intriguing. Starting from a circuit model, we theoretically demonstrate the nonreciprocal behaviour on a transmission line building block creating a strong field asymmetry with a switchable matching stub to enable two distinct working states. After translating to an electromagnetic model, this concept is first proved by simulation and then experimentally verified on a microstrip-line-based diode-integrated metasurface showing nonreciprocal transmission. This printed circuit board design is expected to find various applications in electromagnetic protecting layers, communication systems, microwave isolators and circulators.}


\maketitle
\section{Introduction}\label{sec1}

Reciprocity is a fundamental property of non-magnetic, passive, linear and time-invariant systems, which states that the transmission from the source to the observation point remains the same if their locations are interchanged. In the past decades, with the fast development in acoustics\cite{Nassar20_Review}, electromagnetism\cite{Caloz18_Review,Kord20_Review}, optics\cite{Shoji15_Review,Kutsaev21_Review}, and electronics\cite{Reiskarimian21_Review}, breaking the transmission symmetry becomes a crucial task for various applications, from optical isolation to wireless communication\cite{Xu20_NI_NR,Kong19_NI_NR,Zhang19_STC,Popa14_AE_NR}. In the microwave regime\cite{Pozar11_Fund,Balanis12_Fund}, nonreciprocal metasurfaces raise growing interests because they provide a compact, ultra-thin, lightweight platform to manipulate waves in a unidirectional manner, which enables nonreciprocal wave absorbing, nonreciprocal beam steering, frequency conversion, and consequently can be used to protect sensitive devices from high energy sources\cite{Sievenpiper99_MS,Sievenpiper03_MS,Zhang19_STC,Li20_STM_NR}.

The conventional approach to break reciprocity in electromagnetism relies on ferrite material, such as yttrium iron garnet and iron oxide with other metallic elements, which exhibits the Faraday effect with an external biasing magnetic field parallel to the direction of propagation\cite{Fuller87_Fund,Adam02_Review,Apaydin13_F_AP,Parsa10_F_AP}. Despite that it introduces polarization rotation which might not be preferred in communication systems, ferrites are usually expensive, bulky and incompatible with modern integrated circuits and optical devices.

Recent research proposes to violate the fundamental assumption of the Lorentz reciprocity theorem to break the transmission symmetry by employing spatial-temporal modulation, active elements or nonlinearity\cite{Caloz18_Review,Kord20_Review,Nassar20_Review}. However, such modulation strategy usually requires a complex controlling system which needs an additional power supply, and thus hinders their application\cite{Dinc17_STM_NR,Cardin20_NR_STM,Li20_STM_NR,Yang22_PRA_STM}. In addition, some works adopt active electronic elements, including transistor-based amplifiers, to break the reciprocity\cite{Popa14_AE_NR,Reiskarimian16_AE_NR,Taravati17_AE_NR,Chapman17_AE_NR}. Although modulation is absent, those designs still require a biasing circuit. Nonlinearity, on the other hand, provides an alternative to engineer all-passive nonreciprocal devices. Material with strong Kerr nonliearity has been proven to be a good candidate for nonreciprocal optical devices, including circulators and isolators, if the structure is designed to have field asymmetry when illuminated from opposite directions\cite{Fan12_NI_NR,Jin20_NI_NR,Yang20_NI_NR,Xu20_NI_NR,Mekawy21_NI_NR}.

However, other types of nonlinearity need to be implemented as a counterpart to the electro-optic effect in microwave frequencies to develop an all-passive, magnetic-free nonreciprocal structure. Ref \cite{Mahmoud15_NI_NR} adopts an asymmetrical structure and resonators loaded with an imaginary lumped component, whose capacitance varies with continuous AC signals, to imitate the Kerr effect, such that the resonance frequency shifts differently for waves propagating in opposite directions.

Here, we propose a feasible all-passive nonreciprocal metasurface design without any biasing based on a microstrip line (MSL) with the integration of accessible nonlinear elements, Schottky diodes, to break the transmission symmetry. The simplicity of this model allows fast design in a circuit model by treating the MSL as an ideal transmission line (TL), which makes this method more appealing. The field asymmetry directly originates from the impedance contrast of the MSL, which is later proven to be the determining parameter of the nonreciprocity. The Schottky diodes connect a matching stub to the ground plane, and the field intensity at their location differs significantly for waves propagating in opposite directions. Thus, the matching stub is either open or shorted depending on the incidence direction, which further determines the transmission. In this article, we extend the MSL structure to a three-layered metasurface with low transmission in one direction while allowing wave propagation from the opposite direction, which resembles the diode behaviour. Immune to Faraday rotation, it is potentially applicable in various scenarios including protecting radar or communication systems against high power microwave sources, without using conventional EM circulators and isolators.

\section{Results}\label{sec2}
\subsection{TL Design}\label{subsec2.1}

\begin{figure*}[htbp]%
\centering
\includegraphics[width=1\textwidth]{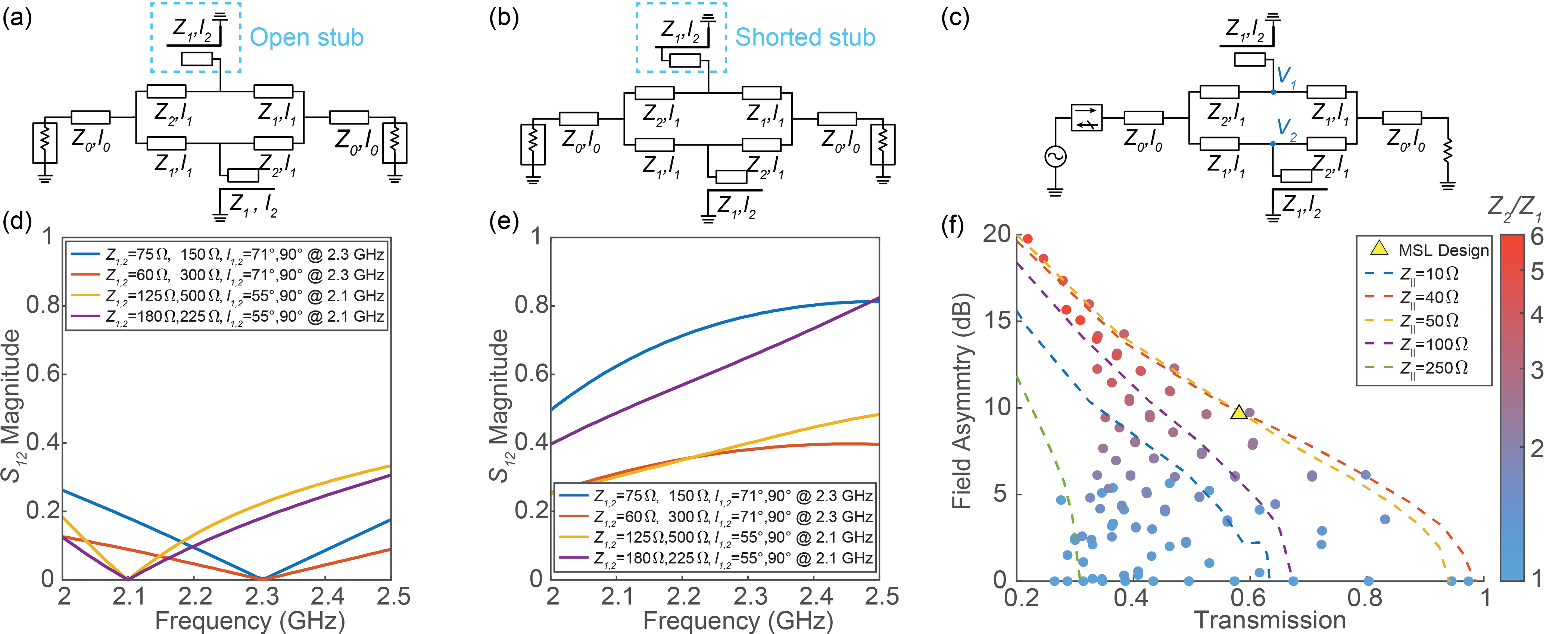}
\caption{Ideal transmission line models. $Z_{0}=50\;\Omega$. (a) OFF state circuit model. (b) ON state circuit model. (c) Transient simulation model in OFF state. (d) Transmission in the OFF state. (e) Transmission in the ON state. (f) Trade-off between field asymmetry and the ON state transmission with $l_{1}=71^{\circ}$, $l_{2}=90^{\circ}$ at $2.3$ GHz.}\label{TL}
\end{figure*}

Nonlinear devices are defined by their power-dependent behaviour. Thus when the field distribution changes between one incident direction and the opposite, placing a nonlinear component at the field asymmetrical location enables the whole system to be nonreciprocal. Take the ideal diode for instance, the ON and OFF working states are determined by the local field intensity and induce two distinct operating states of the system for different transmission/absorption due to the change of its topology. The nonreciprocity is maximized when deploying the diodes at the strongest asymmetry point. Then, a nonreciprocal metasurface can be developed by simply folding this building block and separating the two halves by a ground plane with antennas connected to receive and transmit signals.

The proposed TL model that creates and utilizes a field asymmetry is illustrated in Fig.\ref{TL}. Starting with the ideal circuit model, two different impedances, $Z_{1}$ and $Z_2$, are chosen to create the field asymmetry at the end of the matching stub. Nonreciprocity is then achieved by integrating diodes on the upper stub. As a proof of concept, we simplified the nonlinear power-depended behaviour of the diodes to two static states so that the matching stub is either an open (the OFF state, Fig.\ref{TL} (a)) or a shorted stub (the ON state, Fig.\ref{TL} (b)) depending on the field intensity under different incident directions, while the lower stub is always open. When it is shorted, better matching occurs, resulting in higher transmission. Otherwise, it has lower transmission rate. Since the state of the upper matching stub is determined by the field intensity, a stronger asymmetry in field distribution is crucial for achieving higher nonreciprocity.
 
To analyze the field asymmetry of this TL design, we performed S parameter and transient simulations in Advanced Design System (ADS) to study the static states behaviour. The field asymmetry is defined by the ratio of voltage at two observation points, $V_{1}$ and $V_{2}$ in Fig.\ref{TL} (c) as $20log_{10}(\frac{V_{2}}{V_{1}})$. Although the diode is located at the end of the matching stub, it is reasonable to approximate the ratio of the field intensity at that location by these observation points. The transmission rate for the static states are studied by S parameter simulation as in Fig.\ref{TL} (d) and (e). The length of the matching stub, $L_{2}$, determines the operating frequency which has the highest field asymmetry. The minimum transmission occurs at the frequency when the electrical length of the matching stub is $\lambda/4$ in the OFF state, since the waves are shorted by the matching stub, as shown in Fig.\ref{TL} (d). For the ON state, since one matching stub is shorted, higher transmission is observed as illustrated in (e). $L_{1}$ does not affect the overall performance here since only an ideal TL is involved, but it should be chosen properly in a real electromagnetic design to follow the TL principles. The field asymmetry is dominated by impedance contrast between $Z_{1}$ and $Z_{2}$. By examining different impedance values and contrast for $50\;\Omega\leq Z_{1}\leq Z_{2} \leq 500\;\Omega$, a trade-off between field asymmetry and the ON state transmission is observed in Fig.\ref{TL} (f). All the data in (f) is simulated with $l_{1}=71^{\circ}$, $l_{2}=90^{\circ}$ at $2.3$ GHz.The color in (f) represents the impedance contrast defined by $\frac{Z_{2}}{Z_{1}}$. Warm-colored dots are concentrated on the upper-left corner and the cold-colored ones are scattered at the bottom, demonstrating the trade-off relationship between impedance contrast and field asymmetry. In addition, all data points are gathered in the lower-left half and is bounded by $Z_{\parallel}=50\;\Omega$, where $\frac{1}{Z_{\parallel }}=\frac{1}{Z_{1}}+\frac{1}{Z_{2}}$, which proves the trade-off between field asymmetry and transmission. Higher impedance contrast does create stronger field asymmetry, but the discontinuities and reflections largely undermines the transmission rate. Thus, the upper limit scenario occurs when avoiding the unnecessary reflections due to mismatch. As expected, the field distribution is completely symmetrical if $Z_{1} = Z_{2}$.

\subsection{MSL Design}\label{subsec2.2}

\begin{figure*}[h]%
\centering
\includegraphics[width=1\textwidth]{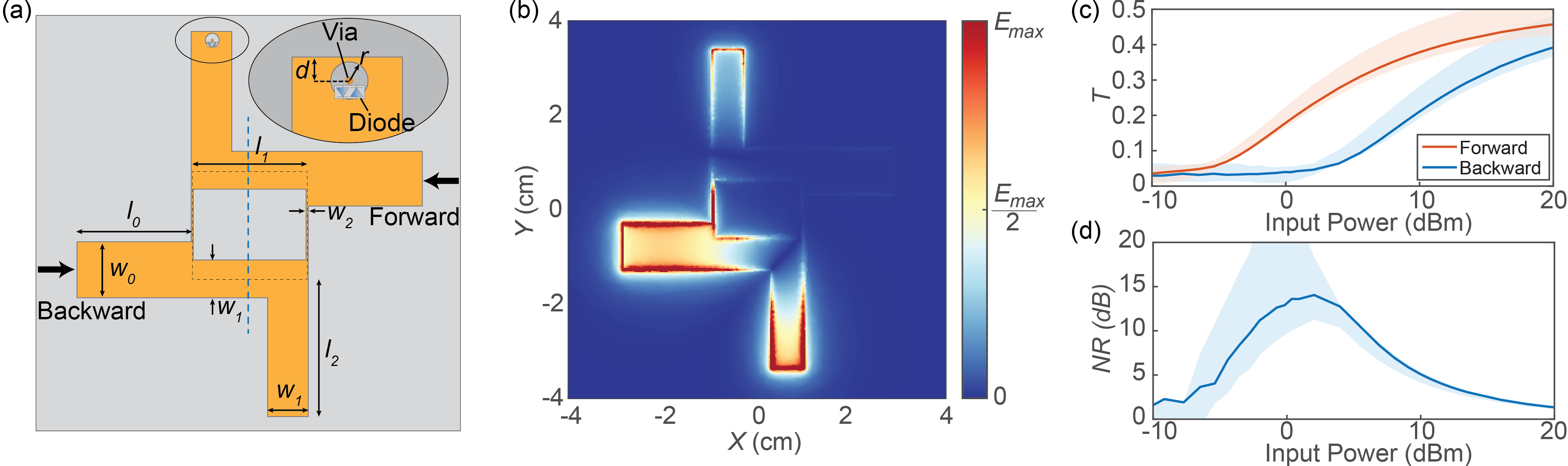}
\caption{Microstrip line model. (a) Illustration of MSL translating from the yellow triangle in Fig.\ref{TL} (f). $w_{0,1,2}$ corresponds to $Z_{0,1,2}=50,66,200\;\Omega$. (b) Field distribution in the OFF state under backward incidence at $2.3$ GHz. (c) The simulated transmission rate under different propagation directions. (d) The simulated nonreciprocity. Solid line: $2.05$ GHz. Shaded area: $2.03$ GHz to $2.15$ GHz.}\label{MSL}
\end{figure*}

To transform the ideal circuit model to a feasible electromagnetic prototype, we adopt a MSL which can be easily fabricated and integrated to printed circuit board (PCB) designs. Fig.\ref{MSL} (a) illustrates a MSL design with a reasonable field asymmetry and transmission rate at the upper limit of the trade-off relation, shown as the yellow triangle in Fig.\ref{TL} (f). The impedance of $Z_{1}$ and $Z_{2}$ is $66\;\Omega$ and $200\;\Omega$, respectively. The length of the matching stub is $l_{2}$ determines the center frequency to be around $2.3$ GHz. $l_{1}$ should be as short as possible to shrink the size, but not to short to still behave as a TL. The thickness of the substrate is chosen such that the width of a $200\;\Omega$ MSL can be precisely manufactured. A pair of oppositely oriented diodes are required to react to sinusoidal waves. Incidence from the right-hand side is defined as forward propagation and excitation from the opposite side is called backward propagation.

The static field distribution in the OFF state is illustrated in Fig.\ref{MSL} (b) under backward excitation. Since this structure has $180^{\circ}$ rotational symmetry, the forward field distribution can be obtained by rotating that of the backward by $180^{\circ}$. That is to say the field asymmetry at the location of the diode can be seen as the ratio between the upper stub and the lower stub under forward or backward incidence. The transmission rate for static forward and backward propagation is $0.65$ and $0.09$ at $2.3$ GHz, and the field asymmetry is $8.3$ dB, which agrees well with the circuit model.

This structure is further studied using EM-circuit co-simulation to verify its power-depended nonreciprocal behaviour with the integration of a real Schottky diode for fast response. Power-related nonreciprocal transmission is observed from $2.03$ GHz to $2.15$ GHz with more than $10$ dB nonreicprocity as shown in Fig.\ref{MSL} (c) and (d). Frequency shift is inevitable due to the integration of real diodes. Here, the transmission ($T$) is the voltage ratio between the received and the transmitted, and the nonreciprocity is defined by 
\begin{equation}
    NR = 20log_{10}(\frac{T_{+}}{T_{-}}),
    \label{eq:NR}
\end{equation}
where $T_{\pm}$ denotes the forward and backward tranmission rate. The incident power is calculated by $\frac{V_{tran.}^2}{2Z_{0}}$, where $V_{tran.}$ is the amplitude of the voltage source. The system is reciprocal under low-power illumination since the diodes are not triggered. As the power increases, the nonreciprocity rises because the diodes are only turned on under forward incidence and it reaches the maximum when the voltage across the diodes of backward excitation is close to the threshold of the diodes. The transmission rate finally converges as the incident power continues increasing, since the threshold of the diodes become negligible compared with the voltage cross the diodes. More than $10$ dB nonreciprocity is achieved between $-2.8$ dBm and $5.6$ dBm incidence for $2.05$ GHz. The static states and power-dependent nonreciprocal performance is further demonstrated through measurement and the results are included in Supplementary Material I, which proves its potentiality as a building block of EM isolators and circulators.

\subsection{Metasurface design and measurement}\label{subsec2.3}
\begin{figure*}[h]%
\centering
\includegraphics[width=1\textwidth]{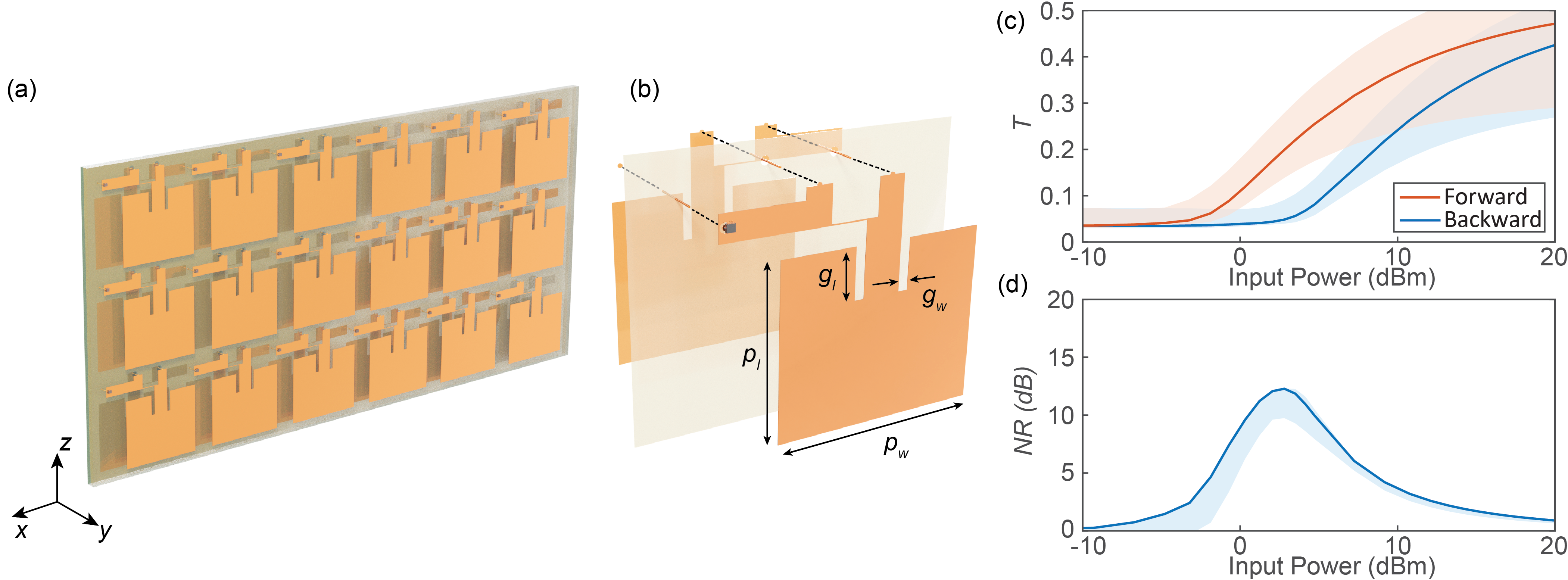}
\caption{Metasurface design. (a) Illustration of a 6 by 3-unit surface. (b) A detailed illustration of the three-layer layout. (c) The simulated transmission rate under different propagation directions. (d) The simulated nonreciprocity. Solid line: 2.21 GHz. Shaded area: 2.18 GHz to 2.25 GHz.}\label{MS}
\end{figure*}

After demonstrating the nonreciprocity of the MSL through simulations and measurements, we extend it to a metasurface, which can be used to protect sensitive devices from high power sources while allowing one-way communication. This is attained by folding the MSL structure and separating the two halves on two layers with a ground plane in between as illustrated in Fig.\ref{MS} (b). These MSL layers are connected by two vias through the holes on the central ground plane. Two patch antennas are attached to the MSL to receive and transmit the EM signals.

The performance of this design is studied using EM-circuit co-simulation on a unit cell with periodic boundaries to reduce the computational cost. Since the excitation is free space waves, the definition of input power is different than that in the MSL simulations, where the incident is calculated by the voltage between the MSL and ground plane. Here, the input refers to the power over the area of one unit cell. Similar to the MSL structure, a nonreciprocal transmission with the same trend is observed as shown in Fig.\ref{MS} (c) and (d). The bandwidth of this unit cell is narrowed down to $2.18$ GHz to $2.25$ GHz due to the patch antenna.

\begin{figure*}[h]%
\includegraphics[width=1\textwidth]{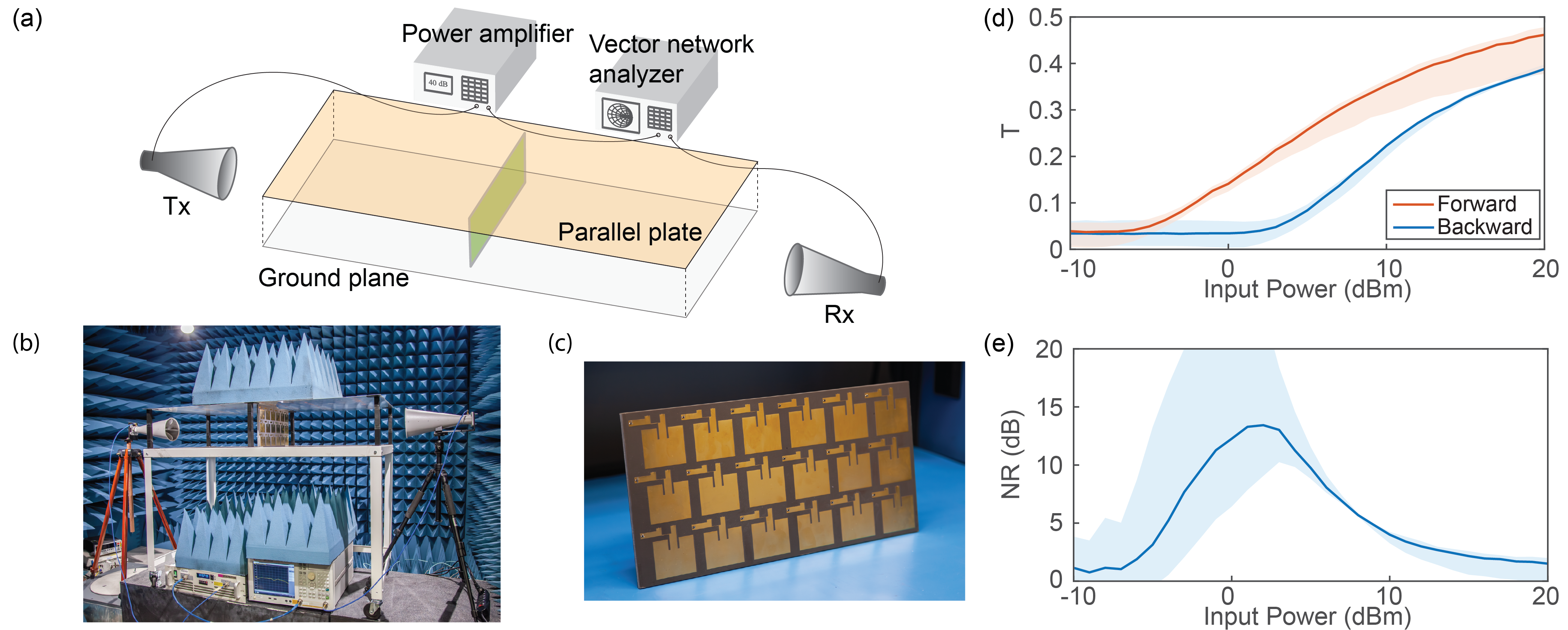}
\caption{Measurement inside a TEM wavegudie to reduce diffraction. (a) Illustration of measurement setup. (b) Picture of measurement setup. (c) Picture of the prototype. (d) The measured transmission rate under different propagation direction. (e) The measured nonreciprocity. Solid line: $2.05$ GHz. Shaded area: $2.02$ GHz to $2.06$ GHz.}\label{Meas.}
\end{figure*}

A $6$ by $3$-unit prototype, as shown in Fig.\ref{MS} (a), is fabricated to measure its nonreciprocity. This experiment is conducted in an anechoic chamber to minimize the effect from the environment. To avoid diffraction and to confine energy propagation, the surface is placed inside a parallel plate TEM waveguide with the same height as illustrated in Fig.\ref{Meas.} (a). Microwave absorbers are placed at the top and the bottom of the waveguide to further reduce diffraction. The incident signal is generated by a vector network analyzer (VNA), amplified by a power amplifier and transmitted by a horn antenna. After propagating through the surface, the signal is received by another horn antenna connected to the VNA. Note that the input power here has the same definition as in Fig.\ref{MS} (c) and (d), referring to the the power incident on each unit cell. Having the setup fixed, the backward transmission is measured by flipping the sample. In addition to the forward and backward transmission scenarios, we also measured the transmission of a blank waveguide (the blank case), representing the total incident power, and the diffraction from the sides with an equal size aluminum sheet (the PEC case). Although the waveguide is wider than the sample, the diffraction measured under the PEC case is $30$ dB less than that of the blank case, which is negligible. Thus, it is reasonable to assume that no diffraction occurs in this measurement. The transmission in the measurement is defined by 
\begin{equation}
    T_{\pm} = \sqrt{\frac{P_{\pm}}{P_{blk}}},
    \label{eq:Meas.T}
\end{equation}
where $P_{\pm/blk}$ is the absolute power received by Rx measured with board under forward/backward propagation and the blank case. The nonreciprocity is still defined by Equ.(\ref{eq:NR}). The measurement results are shown in Fig.\ref{Meas.} (d) and (e), similar to the simulation results in Fig.\ref{MS}, a nonreciprocal transmission behaviour is observed under different incident directions. More than $10$ dB isolation is achieved between $-1.8$ dBm to $4.7$ dBm at $2.05$ GHz.

\section{Discussion}\label{Conclusion}
We proposed a MSL-based all-passive nonreciprocal metasurface by applying nonlinear devices at field asymmetrical locations, and verified its ``diode''-like one-way behavior under opposite incident directions through experiments. Unlike the Kerr effect approach in photonic systems, where the nonlinearity originates from the material, this method is rather general in the microwave regime. Since the field asymmetry comes from the metallic topology, the substrate is free of choice. The option for diodes are also numerous in various packages, material and arrangement. Moreover, it can be scaled to other frequencies, and it is rather easy to modify the input and output impedance by adjusting the width of the MSL. In addition, the working range of power is adjustable using different arrangements of diodes, which is equivalent to adopting diodes with different threshold voltages (Supplementary Material II). Different from the ferrite approach, its immunity to Faraday rotation makes it more appealing for communication systems. Different polarizations can be achieved using other types of antennas, or modifying the feeding point of the current design if polarization conversion is preferred. This lightweight, low-profile and low-cost PCB design is promising in various applications, including EM protecting layers such as radomes and nonreciprocal absorbing material, and EM isolators and circulators by cascading the MSL structure. 
\backmatter

\section{Methods}\label{Methods}
\subsection{Fabrication} The MSL structure is manufactured on a $3.175$ mm thick $8.5$ cm by $8.5$ cm double-sided Duroid Rogers 5880 PCB with coaxial feed. The dimensions are $w_{0,1,2}=9.7,6.63,0.31$ mm, $l_{0,1,2}=1.9,1.9,2.4$ cm, $d=1.4$ mm,$r=1.2$ mm, and the radius of the vias are $10$ mil. The center of the $66\;\Omega$ and $200\;\Omega$ MSL forms a square whose side length is $l_{1}$ shown as the black dotted line in Fig.\ref{MSL}(a). The Schottky diode used in the EM-circuit co-simulation is BAT15-04W. There is a hole on the upper matching stub to place the diodes to connect the stub to the ground plane through a via in the ON state. The metasurface is fabricated on a three-layer $42.3$ cm by $20.5$ cm Duroid Rogers 5880 PCB, both of cores are $3.175$ mm in thickness. The dimensions for the patch antenna, resonating at $2.19$ GHz, are $p_{w}=53.87$ mm, $p_{l}=45$ mm, $g_{w}=2.5$ mm, $g_{l}=13$ mm. Note that the periodicity in $y$ and $z$ direction is $6.8$ cm and $6.6$ cm, respectively, to avoid overlapping of the top and bottom layer for a through via.

\subsection{EM-circuit co simulation} After simulating the EM structure from DC to $8$ GHz, we extract its S parameters and perform transient simulation in circuit with the diodes SPICE model provided by the manufacturer. The metasurface is simulated under periodic boundaries using unit cell.

\subsection{Measurement}
The TEM waveguide is $120$ cm long, $60$ cm wide and its height is $20.3$ cm, which is the same as the prototype. The distance between the edge of the horn antenna (Rx) and the TEM waveguide is $10$ cm, which excites plane wave inside the waveguide. Devices involved: VNA, Keysight E5071C, power amplifier Ophir 5022, Tx and Rx RCDLPHA2G18B. Due to the involvement of an amplifier, instead of S parameters, the VNA is always measuring the absolute power on Rx. To measure under different incident power, the input power of the VNA varies from $-40$ dBm to $-10$ dBm with the amplifier providing a $30$ dB amplification. All the data is averaged by $8$ times, and the devices are controlled by Python. 

\bmhead{Acknowledgments} This work is supported by Office of Naval Research under Grant No. N0014-20-1-2710.

\bmhead{Data availability} All relevant data is available upon request.

\bibliography{sn-article}% common bib file


\begin{thebibliography}{30}
\ifx \bisbn   \undefined \def \bisbn  #1{ISBN #1}\fi
\ifx \binits  \undefined \def \binits#1{#1}\fi
\ifx \bauthor  \undefined \def \bauthor#1{#1}\fi
\ifx \batitle  \undefined \def \batitle#1{#1}\fi
\ifx \bjtitle  \undefined \def \bjtitle#1{#1}\fi
\ifx \bvolume  \undefined \def \bvolume#1{\textbf{#1}}\fi
\ifx \byear  \undefined \def \byear#1{#1}\fi
\ifx \bissue  \undefined \def \bissue#1{#1}\fi
\ifx \bfpage  \undefined \def \bfpage#1{#1}\fi
\ifx \blpage  \undefined \def \blpage #1{#1}\fi
\ifx \burl  \undefined \def \burl#1{\textsf{#1}}\fi
\ifx \doiurl  \undefined \def \doiurl#1{\url{https://doi.org/#1}}\fi
\ifx \betal  \undefined \def \betal{\textit{et al.}}\fi
\ifx \binstitute  \undefined \def \binstitute#1{#1}\fi
\ifx \binstitutionaled  \undefined \def \binstitutionaled#1{#1}\fi
\ifx \bctitle  \undefined \def \bctitle#1{#1}\fi
\ifx \beditor  \undefined \def \beditor#1{#1}\fi
\ifx \bpublisher  \undefined \def \bpublisher#1{#1}\fi
\ifx \bbtitle  \undefined \def \bbtitle#1{#1}\fi
\ifx \bedition  \undefined \def \bedition#1{#1}\fi
\ifx \bseriesno  \undefined \def \bseriesno#1{#1}\fi
\ifx \blocation  \undefined \def \blocation#1{#1}\fi
\ifx \bsertitle  \undefined \def \bsertitle#1{#1}\fi
\ifx \bsnm \undefined \def \bsnm#1{#1}\fi
\ifx \bsuffix \undefined \def \bsuffix#1{#1}\fi
\ifx \bparticle \undefined \def \bparticle#1{#1}\fi
\ifx \barticle \undefined \def \barticle#1{#1}\fi
\bibcommenthead
\ifx \bconfdate \undefined \def \bconfdate #1{#1}\fi
\ifx \botherref \undefined \def \botherref #1{#1}\fi
\ifx \url \undefined \def \url#1{\textsf{#1}}\fi
\ifx \bchapter \undefined \def \bchapter#1{#1}\fi
\ifx \bbook \undefined \def \bbook#1{#1}\fi
\ifx \bcomment \undefined \def \bcomment#1{#1}\fi
\ifx \oauthor \undefined \def \oauthor#1{#1}\fi
\ifx \citeauthoryear \undefined \def \citeauthoryear#1{#1}\fi
\ifx \endbibitem  \undefined \def \endbibitem {}\fi
\ifx \bconflocation  \undefined \def \bconflocation#1{#1}\fi
\ifx \arxivurl  \undefined \def \arxivurl#1{\textsf{#1}}\fi
\csname PreBibitemsHook\endcsname

%%% 1
\bibitem{Nassar20_Review}
\begin{barticle}
\bauthor{\bsnm{Nassar}, \binits{H.}},
\bauthor{\bsnm{Yousefzadeh}, \binits{B.}},
\bauthor{\bsnm{Fleury}, \binits{R.}},
\bauthor{\bsnm{Ruzzene}, \binits{M.}},
\bauthor{\bsnm{Alù}, \binits{A.}},
\bauthor{\bsnm{Daraio}, \binits{C.}},
\bauthor{\bsnm{Norris}, \binits{A.N.}},
\bauthor{\bsnm{Huang}, \binits{G.}},
\bauthor{\bsnm{Haberman}, \binits{M.R.}}:
\batitle{Nonreciprocity in acoustic and elastic materials}.
\bjtitle{Nature Reviews Materials}
\bvolume{5}(\bissue{9}),
\bfpage{667}--\blpage{685}
(\byear{2020})
\end{barticle}
\endbibitem

%%% 2
\bibitem{Caloz18_Review}
\begin{barticle}
\bauthor{\bsnm{Caloz}, \binits{C.}},
\bauthor{\bsnm{Alu}, \binits{A.}},
\bauthor{\bsnm{Tretyakov}, \binits{S.}},
\bauthor{\bsnm{Sounas}, \binits{D.}},
\bauthor{\bsnm{Achouri}, \binits{K.}},
\bauthor{\bsnm{Deck-Léger}, \binits{Z.-L.}}:
\batitle{Electromagnetic nonreciprocity}.
\bjtitle{Physical Review Applied}
\bvolume{10}(\bissue{4}),
\bfpage{047001}
(\byear{2018})
\end{barticle}
\endbibitem

%%% 3
\bibitem{Kord20_Review}
\begin{barticle}
\bauthor{\bsnm{Kord}, \binits{A.}},
\bauthor{\bsnm{Sounas}, \binits{D.L.}},
\bauthor{\bsnm{Alu}, \binits{A.}}:
\batitle{Microwave nonreciprocity}.
\bjtitle{Proceedings of the IEEE}
\bvolume{108}(\bissue{10}),
\bfpage{1728}--\blpage{1758}
(\byear{2020})
\end{barticle}
\endbibitem

%%% 4
\bibitem{Shoji15_Review}
\begin{barticle}
\bauthor{\bsnm{Shoji}, \binits{Y.}},
\bauthor{\bsnm{Miura}, \binits{K.}},
\bauthor{\bsnm{Mizumoto}, \binits{T.}}:
\batitle{Optical nonreciprocal devices based on magneto-optical phase shift in
  silicon photonics}.
\bjtitle{Journal of Optics}
\bvolume{18}(\bissue{1}),
\bfpage{013001}
(\byear{2015})
\end{barticle}
\endbibitem

%%% 5
\bibitem{Kutsaev21_Review}
\begin{barticle}
\bauthor{\bsnm{Kutsaev}, \binits{S.V.}},
\bauthor{\bsnm{Krasnok}, \binits{A.}},
\bauthor{\bsnm{Romanenko}, \binits{S.N.}},
\bauthor{\bsnm{Smirnov}, \binits{A.Y.}},
\bauthor{\bsnm{Taletski}, \binits{K.}},
\bauthor{\bsnm{Yakovlev}, \binits{V.P.}}:
\batitle{Up‐and‐coming advances in optical and microwave nonreciprocity:
  From classical to quantum realm}.
\bjtitle{Advanced Photonics Research}
\bvolume{2}(\bissue{3}),
\bfpage{2000104}
(\byear{2021})
\end{barticle}
\endbibitem

%%% 6
\bibitem{Reiskarimian21_Review}
\begin{barticle}
\bauthor{\bsnm{Reiskarimian}, \binits{N.}}:
\batitle{A review of nonmagnetic nonreciprocal electronic devices: Recent
  advances in nonmagnetic nonreciprocal components}.
\bjtitle{IEEE Solid-State Circuits Magazine}
\bvolume{13}(\bissue{4}),
\bfpage{112}--\blpage{121}
(\byear{2021})
\end{barticle}
\endbibitem

%%% 7
\bibitem{Xu20_NI_NR}
\begin{barticle}
\bauthor{\bsnm{Xu}, \binits{X.-W.}},
\bauthor{\bsnm{Li}, \binits{Y.}},
\bauthor{\bsnm{Li}, \binits{B.}},
\bauthor{\bsnm{Jing}, \binits{H.}},
\bauthor{\bsnm{Chen}, \binits{A.-X.}}:
\batitle{Nonreciprocity via nonlinearity and synthetic magnetism}.
\bjtitle{Physical Review Applied}
\bvolume{13}(\bissue{4}),
\bfpage{044070}
(\byear{2020})
\end{barticle}
\endbibitem

%%% 8
\bibitem{Kong19_NI_NR}
\begin{barticle}
\bauthor{\bsnm{Kong}, \binits{C.}},
\bauthor{\bsnm{Xiong}, \binits{H.}},
\bauthor{\bsnm{Wu}, \binits{Y.}}:
\batitle{Magnon-induced nonreciprocity based on the magnon kerr effect}.
\bjtitle{Physical Review Applied}
\bvolume{12}(\bissue{3}),
\bfpage{034001}
(\byear{2019})
\end{barticle}
\endbibitem

%%% 9
\bibitem{Zhang19_STC}
\begin{barticle}
\bauthor{\bsnm{Zhang}, \binits{L.}},
\bauthor{\bsnm{Chen}, \binits{X.Q.}},
\bauthor{\bsnm{Shao}, \binits{R.W.}},
\bauthor{\bsnm{Dai}, \binits{J.Y.}},
\bauthor{\bsnm{Cheng}, \binits{Q.}},
\bauthor{\bsnm{Castaldi}, \binits{G.}},
\bauthor{\bsnm{Galdi}, \binits{V.}},
\bauthor{\bsnm{Cui}, \binits{T.J.}}:
\batitle{Breaking reciprocity with space‐time‐coding digital metasurfaces}.
\bjtitle{Advanced materials}
\bvolume{31}(\bissue{41}),
\bfpage{1904069}
(\byear{2019})
\end{barticle}
\endbibitem

%%% 10
\bibitem{Popa14_AE_NR}
\begin{barticle}
\bauthor{\bsnm{Popa}, \binits{B.-I.}},
\bauthor{\bsnm{Cummer}, \binits{S.A.}}:
\batitle{Non-reciprocal and highly nonlinear active acoustic metamaterials}.
\bjtitle{Nature communications}
\bvolume{5}(\bissue{1}),
\bfpage{1}--\blpage{5}
(\byear{2014})
\end{barticle}
\endbibitem

%%% 11
\bibitem{Pozar11_Fund}
\begin{bbook}
\bauthor{\bsnm{Pozar}, \binits{D.M.}}:
\bbtitle{Microwave Engineering}.
\bpublisher{John Wiley \& Sons},
\blocation{New York}
(\byear{2011})
\end{bbook}
\endbibitem

%%% 12
\bibitem{Balanis12_Fund}
\begin{bbook}
\bauthor{\bsnm{Balanis}, \binits{C.A.}}:
\bbtitle{Advanced Engineering Electromagnetics}.
\bpublisher{John Wiley \& Sons},
\blocation{NJ}
(\byear{2012})
\end{bbook}
\endbibitem

%%% 13
\bibitem{Sievenpiper99_MS}
\begin{barticle}
\bauthor{\bsnm{Sievenpiper}, \binits{D.}},
\bauthor{\bsnm{Zhang}, \binits{L.}},
\bauthor{\bsnm{Broas}, \binits{R.F.}},
\bauthor{\bsnm{Alexopolous}, \binits{N.G.}},
\bauthor{\bsnm{Yablonovitch}, \binits{E.}}:
\batitle{High-impedance electromagnetic surfaces with a forbidden frequency
  band}.
\bjtitle{IEEE Transactions on Microwave Theory and techniques}
\bvolume{47}(\bissue{11}),
\bfpage{2059}--\blpage{2074}
(\byear{1999})
\end{barticle}
\endbibitem

%%% 14
\bibitem{Sievenpiper03_MS}
\begin{barticle}
\bauthor{\bsnm{Sievenpiper}, \binits{D.F.}},
\bauthor{\bsnm{Schaffner}, \binits{J.H.}},
\bauthor{\bsnm{Song}, \binits{H.J.}},
\bauthor{\bsnm{Loo}, \binits{R.Y.}},
\bauthor{\bsnm{Tangonan}, \binits{G.}}:
\batitle{Two-dimensional beam steering using an electrically tunable impedance
  surface}.
\bjtitle{IEEE Transactions on antennas and propagation}
\bvolume{51}(\bissue{10}),
\bfpage{2713}--\blpage{2722}
(\byear{2003})
\end{barticle}
\endbibitem

%%% 15
\bibitem{Li20_STM_NR}
\begin{barticle}
\bauthor{\bsnm{Li}, \binits{A.}},
\bauthor{\bsnm{Li}, \binits{Y.}},
\bauthor{\bsnm{Long}, \binits{J.}},
\bauthor{\bsnm{Forati}, \binits{E.}},
\bauthor{\bsnm{Du}, \binits{Z.}},
\bauthor{\bsnm{Sievenpiper}, \binits{D.}}:
\batitle{Time-moduated nonreciprocal metasurface absorber for surface waves}.
\bjtitle{Optics Letters}
\bvolume{45}(\bissue{5}),
\bfpage{1212}--\blpage{1215}
(\byear{2020})
\end{barticle}
\endbibitem

%%% 16
\bibitem{Fuller87_Fund}
\begin{bbook}
\bauthor{\bsnm{Fuller}, \binits{A.B.}}:
\bbtitle{Ferrites at Microwave Frequencies}.
\bpublisher{IET},
\blocation{London}
(\byear{1987})
\end{bbook}
\endbibitem

%%% 17
\bibitem{Adam02_Review}
\begin{barticle}
\bauthor{\bsnm{Adam}, \binits{J.D.}},
\bauthor{\bsnm{Davis}, \binits{L.E.}},
\bauthor{\bsnm{Dionne}, \binits{G.F.}},
\bauthor{\bsnm{Schloemann}, \binits{E.F.}},
\bauthor{\bsnm{Stitzer}, \binits{S.N.}}:
\batitle{Ferrite devices and materials}.
\bjtitle{IEEE transactions on microwave theory and techniques}
\bvolume{50}(\bissue{3}),
\bfpage{721}--\blpage{737}
(\byear{2002})
\end{barticle}
\endbibitem

%%% 18
\bibitem{Apaydin13_F_AP}
\begin{barticle}
\bauthor{\bsnm{Apaydin}, \binits{N.}},
\bauthor{\bsnm{Sertel}, \binits{K.}},
\bauthor{\bsnm{Volakis}, \binits{J.L.}}:
\batitle{Nonreciprocal leaky-wave antenna based on coupled microstrip lines on
  a non-uniformly biased ferrite substrate}.
\bjtitle{IEEE transactions on antennas and propagation}
\bvolume{61}(\bissue{7}),
\bfpage{3458}--\blpage{3465}
(\byear{2013})
\end{barticle}
\endbibitem

%%% 19
\bibitem{Parsa10_F_AP}
\begin{barticle}
\bauthor{\bsnm{Parsa}, \binits{A.}},
\bauthor{\bsnm{Kodera}, \binits{T.}},
\bauthor{\bsnm{Caloz}, \binits{C.}}:
\batitle{Ferrite based non-reciprocal radome, generalized scattering matrix
  analysis and experimental demonstration}.
\bjtitle{IEEE transactions on antennas and propagation}
\bvolume{59}(\bissue{3}),
\bfpage{810}--\blpage{817}
(\byear{2010})
\end{barticle}
\endbibitem

%%% 20
\bibitem{Dinc17_STM_NR}
\begin{barticle}
\bauthor{\bsnm{Dinc}, \binits{T.}},
\bauthor{\bsnm{Tymchenko}, \binits{M.}},
\bauthor{\bsnm{Nagulu}, \binits{A.}},
\bauthor{\bsnm{Sounas}, \binits{D.}},
\bauthor{\bsnm{Alu}, \binits{A.}},
\bauthor{\bsnm{Krishnaswamy}, \binits{H.}}:
\batitle{Synchronized conductivity modulation to realize broadband lossless
  magnetic-free non-reciprocity}.
\bjtitle{Nature communications}
\bvolume{8}(\bissue{1}),
\bfpage{1}--\blpage{9}
(\byear{2017})
\end{barticle}
\endbibitem

%%% 21
\bibitem{Cardin20_NR_STM}
\begin{barticle}
\bauthor{\bsnm{Cardin}, \binits{A.E.}},
\bauthor{\bsnm{Silva}, \binits{S.R.}},
\bauthor{\bsnm{Vardeny}, \binits{S.R.}},
\bauthor{\bsnm{Padilla}, \binits{W.J.}},
\bauthor{\bsnm{Saxena}, \binits{A.}},
\bauthor{\bsnm{Taylor}, \binits{A.J.}},
\bauthor{\bsnm{Kort-Kamp}, \binits{W.J.}},
\bauthor{\bsnm{Chen}, \binits{H.-T.}},
\bauthor{\bsnm{Dalvit}, \binits{D.A.}},
\bauthor{\bsnm{Azad}, \binits{A.K.}}:
\batitle{Surface-wave-assisted nonreciprocity in spatio-temporally modulated
  metasurfaces}.
\bjtitle{Nature communications}
\bvolume{11}(\bissue{1}),
\bfpage{1}--\blpage{9}
(\byear{2020})
\end{barticle}
\endbibitem

%%% 22
\bibitem{Yang22_PRA_STM}
\begin{barticle}
\bauthor{\bsnm{Yang}, \binits{X.}},
\bauthor{\bsnm{Wen}, \binits{E.}},
\bauthor{\bsnm{Sievenpiper}, \binits{D.F.}}:
\batitle{Broadband time-modulated absorber beyond the bode-fano limit for short
  pulses by energy trapping}.
\bjtitle{Physical Review Applied}
\bvolume{17}(\bissue{4}),
\bfpage{044003}
(\byear{2022})
\end{barticle}
\endbibitem

%%% 23
\bibitem{Reiskarimian16_AE_NR}
\begin{barticle}
\bauthor{\bsnm{Reiskarimian}, \binits{N.}},
\bauthor{\bsnm{Krishnaswamy}, \binits{H.}}:
\batitle{Magnetic-free non-reciprocity based on staggered commutation}.
\bjtitle{Nature communications}
\bvolume{7}(\bissue{1}),
\bfpage{1}--\blpage{10}
(\byear{2016})
\end{barticle}
\endbibitem

%%% 24
\bibitem{Taravati17_AE_NR}
\begin{barticle}
\bauthor{\bsnm{Taravati}, \binits{S.}},
\bauthor{\bsnm{Khan}, \binits{B.A.}},
\bauthor{\bsnm{Gupta}, \binits{S.}},
\bauthor{\bsnm{Achouri}, \binits{K.}},
\bauthor{\bsnm{Caloz}, \binits{C.}}:
\batitle{Nonreciprocal nongyrotropic magnetless metasurface}.
\bjtitle{IEEE Transactions on Antennas and Propagation}
\bvolume{65}(\bissue{7}),
\bfpage{3589}--\blpage{3597}
(\byear{2017})
\end{barticle}
\endbibitem

%%% 25
\bibitem{Chapman17_AE_NR}
\begin{barticle}
\bauthor{\bsnm{Chapman}, \binits{B.J.}},
\bauthor{\bsnm{Rosenthal}, \binits{E.I.}},
\bauthor{\bsnm{Kerckhoff}, \binits{J.}},
\bauthor{\bsnm{Moores}, \binits{B.A.}},
\bauthor{\bsnm{Vale}, \binits{L.R.}},
\bauthor{\bsnm{Mates}, \binits{J.}},
\bauthor{\bsnm{Hilton}, \binits{G.C.}},
\bauthor{\bsnm{Lalumiere}, \binits{K.}},
\bauthor{\bsnm{Blais}, \binits{A.}},
\bauthor{\bsnm{Lehnert}, \binits{K.}}:
\batitle{Widely tunable on-chip microwave circulator for superconducting
  quantum circuits}.
\bjtitle{Physical Review X}
\bvolume{7}(\bissue{4}),
\bfpage{041043}
(\byear{2017})
\end{barticle}
\endbibitem

%%% 26
\bibitem{Fan12_NI_NR}
\begin{barticle}
\bauthor{\bsnm{Fan}, \binits{L.}},
\bauthor{\bsnm{Wang}, \binits{J.}},
\bauthor{\bsnm{Varghese}, \binits{L.T.}},
\bauthor{\bsnm{Shen}, \binits{H.}},
\bauthor{\bsnm{Niu}, \binits{B.}},
\bauthor{\bsnm{Xuan}, \binits{Y.}},
\bauthor{\bsnm{Weiner}, \binits{A.M.}},
\bauthor{\bsnm{Qi}, \binits{M.}}:
\batitle{An all-silicon passive optical diode}.
\bjtitle{Science}
\bvolume{335}(\bissue{6067}),
\bfpage{447}--\blpage{450}
(\byear{2012})
\end{barticle}
\endbibitem

%%% 27
\bibitem{Jin20_NI_NR}
\begin{barticle}
\bauthor{\bsnm{Jin}, \binits{B.}},
\bauthor{\bsnm{Argyropoulos}, \binits{C.}}:
\batitle{Self-induced passive nonreciprocal transmission by nonlinear bifacial
  dielectric metasurfaces}.
\bjtitle{Physical Review Applied}
\bvolume{13}(\bissue{5}),
\bfpage{054056}
(\byear{2020})
\end{barticle}
\endbibitem

%%% 28
\bibitem{Yang20_NI_NR}
\begin{barticle}
\bauthor{\bsnm{Yang}, \binits{K.Y.}},
\bauthor{\bsnm{Skarda}, \binits{J.}},
\bauthor{\bsnm{Cotrufo}, \binits{M.}},
\bauthor{\bsnm{Dutt}, \binits{A.}},
\bauthor{\bsnm{Ahn}, \binits{G.H.}},
\bauthor{\bsnm{Sawaby}, \binits{M.}},
\bauthor{\bsnm{Vercruysse}, \binits{D.}},
\bauthor{\bsnm{Arbabian}, \binits{A.}},
\bauthor{\bsnm{Fan}, \binits{S.}},
\bauthor{\bsnm{Alù}, \binits{A.}}:
\batitle{Inverse-designed non-reciprocal pulse router for chip-based lidar}.
\bjtitle{Nature Photonics}
\bvolume{14}(\bissue{6}),
\bfpage{369}--\blpage{374}
(\byear{2020})
\end{barticle}
\endbibitem

%%% 29
\bibitem{Mekawy21_NI_NR}
\begin{botherref}
\oauthor{\bsnm{Mekawy}, \binits{A.}},
\oauthor{\bsnm{Sounas}, \binits{D.L.}},
\oauthor{\bsnm{Alù}, \binits{A.}}:
Free-space nonreciprocal transmission based on nonlinear coupled fano
  metasurfaces.
In: Photonics,
vol. 8,
p. 139.
MDPI
\end{botherref}
\endbibitem

%%% 30
\bibitem{Mahmoud15_NI_NR}
\begin{barticle}
\bauthor{\bsnm{Mahmoud}, \binits{A.M.}},
\bauthor{\bsnm{Davoyan}, \binits{A.R.}},
\bauthor{\bsnm{Engheta}, \binits{N.}}:
\batitle{All-passive nonreciprocal metastructure}.
\bjtitle{Nature communications}
\bvolume{6}(\bissue{1}),
\bfpage{1}--\blpage{7}
(\byear{2015})
\end{barticle}
\endbibitem

\end{thebibliography}

\end{document}